Disk-Integrated Thermal Properties of Ceres Measured at Millimeter Wavelengths

Jian-Yang Li[1] (李荐扬), Arielle Moullet[2], Timothy N. Titus[3], Henry H. Hsieh[1], Mark V. Sykes[1]

[1] Planetary Science Institute, 1700 E. Ft. Lowell Rd., Tucson AZ, 85719
[2] SOFIA/USRA, Moffett Field CA
[3] US Geological Survey, Flagstaff AZ

Abstract

We observed Ceres at three epochs in 2015 November and 2017 September and October with ALMA 12-meter array and in 2017 October with the ALMA Compact Array (ACA), all at ~265 GHz continuum (wavelengths of ~1.1 mm) to map the temperatures of Ceres over a full rotation at each epoch. We also used 2017 October ACA observations to search for HCN. The disk-averaged brightness temperature of Ceres is measured to be between 170 K and 180 K during our 2017 observations. The rotational lightcurve of Ceres shows a double peaked shape with an amplitude of about 4%. Our HCN search returns a negative result with an upper limit production rate of ~$2\times10^{24}$ molecules $s^{-1}$, assuming globally uniform production and a Haser model. A thermophysical model suggests that Ceres's top layer has higher dielectric absorption than lunar-like materials at a wavelength of 1 mm. However, previous observations showed that the dielectric absorption of Ceres decreases towards longer wavelengths. Such distinct dielectric properties might be related to the hydrated phyllosilicate composition of Ceres and possibly abundant µm-sized grains on its surface. The thermal inertia of Ceres is constrained by our modeling as likely being between 40 and 160 tiu, much higher than previous measurements at infrared wavelengths. Modeling also suggests that Ceres's lightcurve is likely dominated by spatial variations in its physical or compositional properties that cause changes in Ceres's observed thermal properties and dielectric absorption as it rotates.

1. Introduction

Recent evidence from investigations of the mineralogy, composition and geology of Ceres performed by the Dawn mission (e.g., De Sanctis et al. 2016, Ruesch et al. 2016, etc.) places Ceres into the category of "candidate ocean worlds" (Hendrix et al. 2019). The dwarf planet's salt-rich subsurface material and possible cryovolcanic and geothermal activity suggests the existence of briny liquids at depth (Ruesch et al. 2018, Sori et al. 2018, Scully et al. 2019). Surface geomorphology analyses indicate that Ceres's outer shell is composed of a mixture of ice, rock, salts, and/or clathrates (Sizemore et al. 2017, Hiesinger et al. 2016, Bland et al. 2016). Hydrogen abundance measurements of Ceres are consistent with abundant water ice at depths of meters (Prettyman et al. 2017). The distribution of additional ice found towards high latitude regions is consistent with the thermal modeling prediction that Ceres could have preserved shallow subsurface water ice at the present time (Schorghofer 2008, 2016, Titus et al. 2015). Exposed water ice in isolated patches with areas of up to a few $km^2$ have been identified in a number of fresh craters on Ceres (Combe et al. 2016, 2019) and inside the permanently shadowed craters in the polar regions (Platz et al. 2016, Schorghofer et al. 2016). One possible detection and one relatively certain detection of water vaporization from Ceres (A'Hearn and Feldman 1992, Küppers et al. 2014) have revealed a possibly active world.

Given the above considerations, understanding the thermal conditions on Ceres is especially important for constraining the present-day distribution of subsurface water ice on the body, as well as the history of water in the evolution of Ceres. As such, we observed Ceres at mm wavelengths with the Atacama Large Millimeter/submillimeter Array (ALMA) to ascertain its temperature and subsurface thermal properties. We also tuned a sideband of the receiver to center at the HCN J=3-2 transition frequency to search for this gaseous species around Ceres. HCN is a common gas species found in cometary comae that is generally considered to originate from the nucleus (e.g., Cordiner et al. 2014). The extended array configuration allowed us to map the distribution of thermal emission over the disk of Ceres and track the rotation. In this article, we report the results of a disk-integrated photometric analysis of Ceres, thermal modeling results, and the results of our HCN search.

Ceres has been observed at radio wavelengths in the past to measure its brightness temperature and rotational lightcurve. Webster et al. (1988) and Webster and Johnston (1989) discussed the decrease of brightness temperatures measured from mm to dm (decimeter) wavelengths, and suggested that surface dielectric properties have a strong effect on microwave observations. Redman et al. (1998) performed a detailed analysis of the different spectral energy distributions (SEDs) of a number of asteroids including Ceres from thermal infrared to dm wavelengths, and showed that those SEDs are related to the mineralogical composition of asteroid surfaces. Meanwhile, Keihm et al. (2013) performed a detailed self-consistent thermophysical and radiative transfer modeling to reconcile the SEDs of four large asteroids from thermal infrared to dm wavelengths and confirmed the low thermal inertia of the top layers of those asteroids. These previous works inform our basic approach to interpreting our observations of Ceres.

On the other hand, inconsistent results have been reported for the rotational lightcurve of Ceres at mm wavelengths. Altenhoff et al. (1996) and Moullet et al. (2010) both reported a lightcurve amplitude of ~3% at 250 GHz and 235 GHz, respectively, while Chamberlain et al. (2009) reported a 50% peak-to-peak amplitude at 345 GHz.

Previous observations in the thermal infrared suggested that the top layer of Ceres's regolith has low thermal inertia. Spencer (1990) derived a thermal inertia of 15 J $m^{-2}$ $s^{-0.5}$ $K^{-1}$ (thermal inertia units, hereafter "tiu") for Ceres. Saint-Pé et al. (1993) reported a value of 38±14 tiu based on ground-based adaptive optics observations of Ceres in the M-band (3.55 – 4.15 µm). Recent observations by Dawn's near-infrared mapping spectrometer at 3.5 – 5 µm have revealed an area of high thermal inertia of (130 – 140 tiu) in the Haulani crater region whereas, for reference, surrounding surface material has a comparatively low overall thermal inertia of up to 60 tiu (Rognini et al. 2019).

We describe our observations in Section 2, and report the results of these observations in Section 3. Thermophysical and radiative transfer modeling is presented in Section 4. In Section 5, we discuss the implications of our observational results and modeling for the thermal and dielectric properties of Ceres's surface. The last section summarizes the conclusions.

## 2. Observations

### 2.1. Observations

We observed Ceres at three epochs in ALMA Cycles 3, 4, and 5 (Table 1), each covering about one full rotation (9 hours) of the object. In order to minimize high airmass observations at low elevation, and to achieve the highest possible sky resolution to support the mapping, we only observed Ceres when it was close to transit with respect to the local meridian. To accomplish this, we divided the observations in each epoch into three separate segments, each about three hours long, to be executed over three days as close to each other as possible as weather permitted, but also timed to cover different longitudinal ranges on Ceres to provide a full rotational coverage. Dates and times of our observations and the corresponding observing geometries and longitudinal coverage, as well as the measured total flux densities and corresponding brightness temperatures are listed in Table 1.

The 12-m array was used for all three epochs, with a primary objective of high-resolution mapping, which will be addressed in a later paper. At the time of our observations the array was in extended configurations with the longest baselines reaching up to 15 km with 40 antennas at a minimum, allowing one to resolve Ceres's disk in 10 – 20 beams across.

In Cycle 5, we also used the Atacama Compact Array (ACA) in addition to the 12-m array. ACA observations, with a spatial resolution of 5", did not spatially resolve Ceres. The purposes of the ACA data were to better measure Ceres's total flux and rotational lightcurve, specifically avoiding issues of missing flux due to Ceres being over-resolved, which can affect our 12-m observations, and to search for HCN gas around Ceres.

All 12-m array observations were performed in Band 6 with a spectral setup optimized for continuum measurement sensitivity, using four 2 GHz-wide spectral windows tuned to 256, 258, 272, and 274 GHz for a total of 7.5 GHz of effective bandwidth. ACA observations were also obtained in Band 6 but with slightly different tuning such that one spectral window was centered at the rest frequency of HCN J=3-2 transition of 265.886 GHz with a high spectral resolution of 244 kHz (275 m/s at the line frequency), and three 2 GHz-wide spectral windows tuned at 252.5, 254.5, 269.5 GHz were dedicated to continuum integration.

### 2.2. Data reduction and flux calibration

All data was calibrated using ALMA's automated science pipeline package, CASA (McMullin et al., 2007). Calibration steps included the flagging of outlier data, correction of the spectral response (bandpass) using a bright reference quasar, correction of temporal gain variations (in phase and amplitude) using a nearby quasar, as well as adjustment of the absolute flux scale using reference flux calibration sources chosen automatically by the telescope operating system.

Achieving reliable absolute flux calibration is paramount for our scientific purpose. To accomplish that, we need to use bright and well monitored sources as anchors, and if possible, the same source for observations obtained close in time. The default flux calibrator used by the automated pipeline is not necessarily the best source recorded in the data for achieving consistent flux calibration across observations, so we refined the absolute flux scale using bandpass or phase calibrators instead. Specifically, for our Cycle 3 data in 2015, quasar J1924-2914 was selected as the most reliable calibrator for all observations, and for our data from Cycles 4 and 5 (both 12-m and ACA) in 2017, quasar J0854+2006 was the best available calibrator. We estimate a 5% absolute calibration uncertainty corresponding to the uncertainty on the quasar flux model.

After all these steps are performed, the data consist of calibrated visibilities – complex measurements corresponding to samples of the Fourier transform of the sky brightness distribution. Visibilities can be directly analyzed to retrieve fluxes, for example by averaging amplitude values over a section of time (for ACA observations, which do not spatially resolve the source), or performing visibility fitting using a disk model.

For 12-m data, one can image visibilities by performing inverse Fourier transforms and deconvolution using cleaning options available in CASA. Given the high relative spatial resolution compared to the size of the source, the best results were obtained by subtracting a uniform disk model, cleaning the residuals, and then adding the initial model to the cleaned components. With the source being bright and having a well-defined shape, phase self-calibration based on imaging results was also applied successfully to improve corrections of short-term phase gain variations. The obtained images can then be directly analyzed.

## 3. Results

### 3.1. Rotational lightcurve

We measured the total flux density of Ceres by integrating over all pixels in the images assembled from the 12-m array data integrated over a short amount of time (5 to 15 min depending on the Fourier plane coverage), in order to achieve good temporal sampling. Sky background is not a concern in interferometric data, as large spatial features are filtered out by the absence of short spacings in the measured Fourier plane. Ceres had apparent angular diameter of 0.43" to 0.48" in our data. We used circular apertures with diameters of 0.6" to 0.8" centered on Ceres's disk to make integrated photometry measurements. Our measured flux densities are independent of the aperture size and slight offsets from the aperture center. In addition to aperture photometry, we also used Fourier transform visibility fitting to measure the total flux density of Ceres received by the antennae and found similar results. The average frequency of the 12-m data is 265 GHz.

For our ACA observations, we measured fluxes directly from the calibrated visibilities by averaging the visibility amplitudes measurements with the same time stamp over all antenna pairs, all spectral windows and both polarizations. The average frequency of ACA flux data is 259.39 GHz.

Rotational lightcurves from all four data sets that we obtained (three epochs of 12-m data and ACA data from Cycle 5) are shown in Figure 1a. Narrow longitudinal gaps exist in each lightcurve, but overall, the entire rotation of Ceres is covered. We note that lightcurves derived from 12-m data from Cycle 3 and Cycle 5 are noisier than those derived from Cycle 4 and Cycle 5 ACA data. Specifically, the Cycle 3 lightcurve dips in the 2015 November 12 data to as low as 60% of the flux density level on other days (see Table 1). The Cycle 5 12-m lightcurve also has dips in the 2017 October 15 data down to about 80% of the flux density level on other days. On the other hand, the lightcurves from Cycle 4 12-m data and Cycle 5 ACA data are consistent in both the levels of flux density and shapes over the rotation of Ceres (Fig. 1b).

In order to understand the reliability of our lightcurves, we considered various sources of uncertainties in our flux measurements, including absolute radiometric calibration, loss of extended flux by interferometry array, varying weather conditions during observations, and photon and electronic noises. First, as discussed in Section 2.2, the absolute radiometric calibration uncertainty is about 5% and is systematic if using the same reference quasar (all data from Cycles 4 and 5). Second, we assessed the potential flux loss in our data by simulating our observations using a model disk of the size of Ceres sampled at the same Fourier plane coverages as the data. Our tests suggested no issues recovering the total flux of Ceres from our data. We do not expect any flux loss from Cycle 5 ACA data because the spatial resolution (~5") is much larger than the angular size of Ceres (~0.5"). Third, unstable weather conditions during an observation, or changing weather from one day to another introduce statistical uncertainties in the lightcurve. Indeed, on 2015 November 12 and 2017 October 15, the phase RMS, which characterizes the visibility phase stability measured on the phase calibrator over relatively short time-scales, was 40º - 60º and 60º - 90º per antenna, respectively, even after short-term path length variation corrections derived from the water vapor radiometer. High phase RMS causes coherence loss and flux loss, especially on the antennas that are farthest apart, and flux loss significantly varied in time and cannot be recovered. The lightcurves including those two dates show clear flux drops (Fig. 1a). Therefore, bad weather conditions rendered the lightcurves derived from Cycle 3 and Cycle 5 12-m data unreliable for further analysis. In contrast, Cycle 4 data have much lower phase RMS with a median of ~15º per antenna, and the corresponding lightcurves appear to be much less noisy. Finally, we estimated the sensitivity of our data. Thermal noise (including noise due to atmospheric opacity and to electronics), estimated using the ALMA Sensitivity Calculator[1], is on the order of μJy and thus negligible. The sensitivity in the maps is instead dominated by the ability to reconstruct the brightness distribution of a large and bright source, i.e., imaging dynamics. There may be residuals from the deconvolution of the dirty beam in the image. This residual is evident from measuring the RMS in the images, which ranges from 0.07 to 0.6 mJy per beam, dominating over the thermal noise. A conservative estimate of the signal-to-noise ratio of our measurements is hence >120.

Based on the above uncertainty analysis, the lightcurves from Cycle 4 data and Cycle 5 ACA data are the most reliable, as confirmed by their consistent flux levels and shapes with one another over the rotation of Ceres (Fig. 1b). The excellent agreement between these two lightcurves strongly suggests that the features in the lightcurve are real, and we will therefore rely only on these two datasets to characterize Ceres's lightcurve at mm wavelengths. Based on Fig. 1b, the maximum rotational variation of the observed Ceres flux density is 4% (peak-to-peak). The overall flux density level is relatively low in the eastern hemisphere (longitudes of 0º – 180º) than in the western hemisphere (longitudes of 180º – 360º). Our data reveal a double-peaked lightcurve of Ceres at a wavelength of 1-mm, with the two minima near longitudes of 30º and 190º and two maxima near longitudes of 120º and 330º. Both the lightcurve amplitude and the overall shape and phasing are consistent with previous observations at similar frequencies (Altenhoff et al. 1996, Moullet et al. 2010), although the detailed shape of those previously measured lightcurves was obscured by noise.

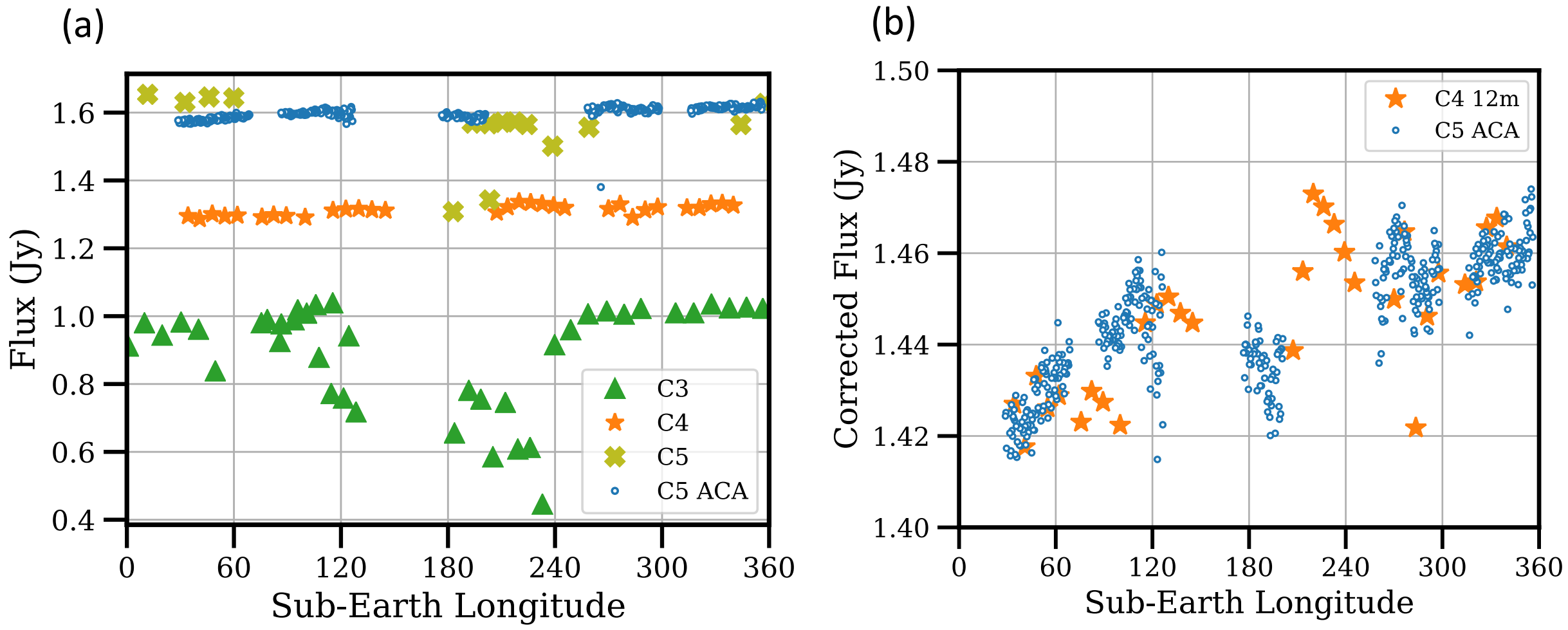

Fig. 1. Lightcurves of Ceres from our data (a), and comparison between the lightcurves from Cycle 4 12-m data and Cycle 5 ACA data (b). Within each epoch, the fluxes are corrected to the mean geocentric distance. In panel (b), the fluxes are further corrected to the mean geocentric distance of Cycle 4 12-m data and Cycle 5 ACA data. No error bars are shown in the plots because, as discussed in Section 3.1, thermal and imaging noises are <1%, and the statistical uncertainties in the lightcurves are dominated by weather conditions and hard to quantify. On the other hand, the point-to-point scatter in the lightcurve provide a reasonable estimate of the statistical errors of our flux measurements.

[1] https://almascience.nrao.edu/proposing/sensitivity-calculator, accessed on March 16, 2020.

3.2. Brightness temperature variations

We calculated brightness temperatures for Ceres based on total flux density measurements using the Rayleigh-Jeans approximation (Table 1, Fig. 2). Effective emissivity can be calculated from the disk-averaged temperature distribution based on a standard thermal model (STM, Lebofsky et al. 1986) with no infrared beaming parameter. For the observing geometry of our Cycles 4 and 5 data, the expected brightness temperature of Ceres is 216 K, resulting in an effective emissivity of 0.8, consistent with previous observations at mm wavelengths (e.g., Redman et al. 1998).

While brightness temperature measurements from Cycles 4 and 5 data are all consistent with historical measurements (Table 2), values derived from Cycle 3 data are much lower than those from all other epochs and historical values except for the one from Altenhoff et al. (1994). In order to confirm the total flux density measurement from our Cycle 3 data, we searched the ALMA archive for ACA data of Ceres acquired at times close to our Cycle 3 observations and found four sets of data. We measured the total flux densities of Ceres from those data following the same approach as for our own ACA data as described in Section 2.2, and calculated the corresponding brightness temperatures. Although analysis of archival ACA data results in a range of brightness temperature values, they are all much higher than our derived Cycle 3 brightness temperatures. Therefore, it is likely that the reconstruction of Ceres interferometry images from our Cycle 3 data did not recover all the flux from Ceres. However, it is still quite puzzling that the brightness temperature reported by Altenhoff et al. (1994) based on single-dish data is similar to our Cycle 3 measurements, although their error bar is quite large. In addition, two other measurements from previous observations at similar true anomalies (Ulich et al. 1984, Webster et al. 1988), both based on single-dish data as well, show a similar range of brightness temperatures as the values we derive from archival ALMA data. At three other true anomalies (0º, 220º, and 310º), repeated measurements resulted in similar brightness temperatures from both interferometer data (Moullet et al. 2010) and single-dish data (all other historical data). Therefore, the possibility of temporal variations in Ceres's brightness temperature at a wavelength of 1 mm near the true anomalies of 150º - 180º still cannot be fully rejected.

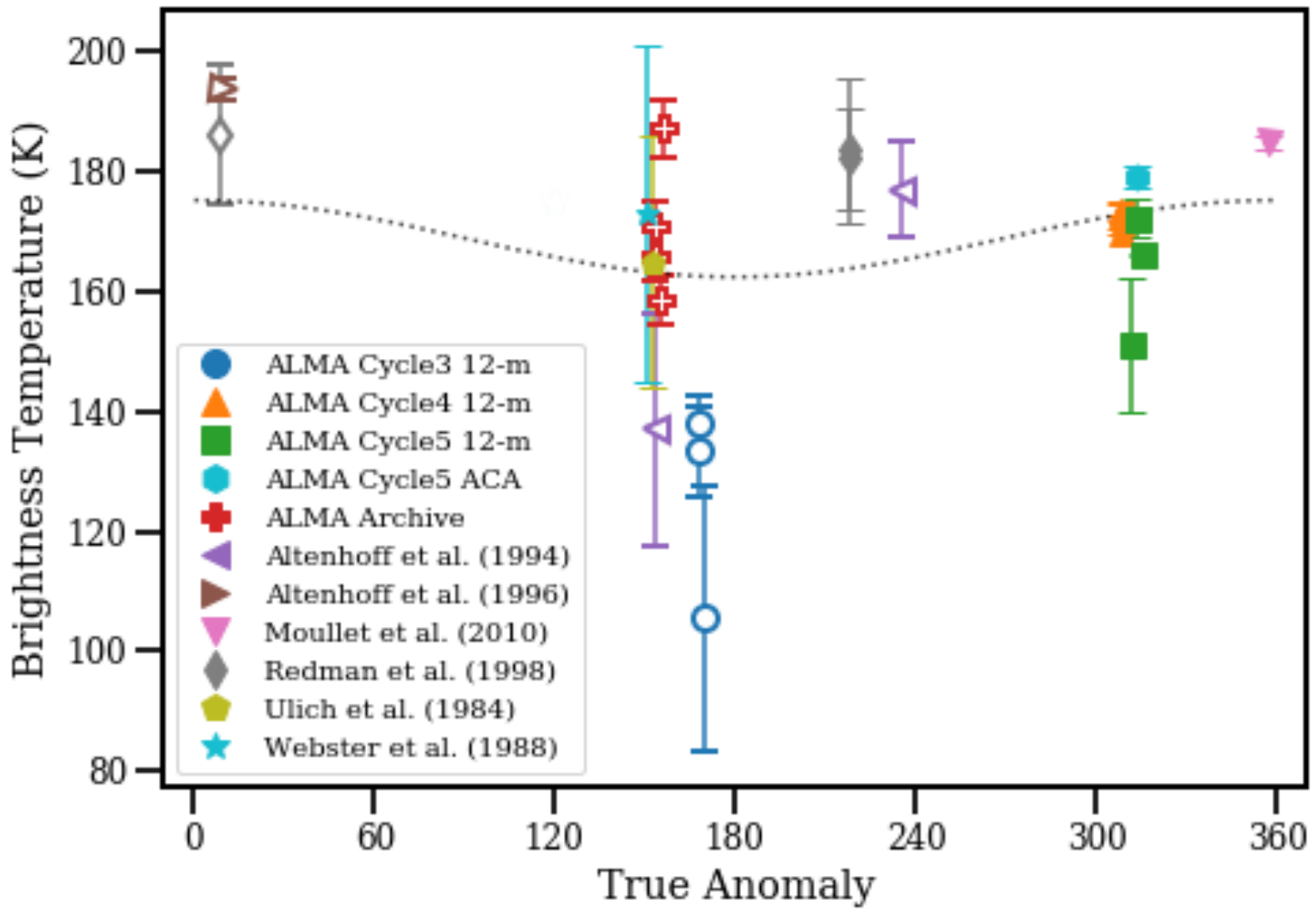

Fig. 2. Brightness temperature measurements of Ceres from our data and from historical observations. Open symbols are pre-opposition measurements with the sub-Earth point in local afternoon, and filled symbols are post-opposition measurements with the sub-Earth point in local morning. The dotted line marks the expected annual temperature curve based on an STM and an assumed effective emissivity of 0.8 with no infrared beaming. The error bars for our measurements represent the standard deviations of data in each day. The error bars for measurements from archival ACA data are estimated to be 5%. The error bars for historical values are quoted from the literature. Perihelion is at a true anomaly of 0º, and aphelion is at a true anomaly of 180º.

Combining all brightness temperature measurements and ignoring the large range of values between true anomalies of 150º and 180º for now, we find the possibility of a seasonal trend in our data. The range of annual brightness temperature variations is about 20 K. The maximum temperature is around a true anomaly of 0º to 60º, consistent with Ceres reaching its minimum heliocentric distance at perihelion and a large thermal delay. The minimum temperature appears to be near a true anomaly of 240º - 300º, which represents an even larger phase lag from aphelion compared to that for the perihelion. Such asymmetric thermal lags with respect to perihelion and aphelion, as well as possible temporal variations of the thermal brightness temperature at a true anomaly of 150º - 180º could be indicators of unusual thermal behavior at the annual thermal skin depth, which is on the order of decimeters on Ceres.

### 3.3. Search for HCN

We used ACA spectral data collected in 2017 October to search for HCN (Table 1). Data from 2017 October 15 are noisy due to bad weather and are excluded from this analysis. Spectral data from 2017 October 18, 19, and 20 are averaged over all observing times, all antenna pairs, and two polarization settings. To increase the signal-to-noise ratio of the spectrum, we further smoothed the spectrum by binning the data to a spectral resolution of 300 kHz, or an equivalent velocity resolution of 0.34 km/s, which is about 60% of the expected average thermal velocity of about 0.5 km/s for Ceres exosphere at the average surface temperature of about 170 K. The 1-$\sigma$ noise level is about 0.018 Jy in the smoothed spectrum.

No HCN J=3-2 emission line is detected in the smoothed spectrum (Fig. 3). We estimated the upper limit production rate of HCN using the Planetary Spectrum Generator[2] (PSG, Villanueva et al. 2018) by finding the production rate that has the flux density at the peak of the J=3-2 emission line comparable to our 3-$\sigma$ noise level of 0.054 Jy. No rotational temperature measurement of HCN is available at comparable heliocentric distance as Ceres. Therefore, we used the empirical relationship $T_{rot} = 60/r_h$ derived from observations of cometary comae, where $T_{rot}$ is in K, and $r_h$ is the heliocentric distance of target in au (Villanueva et al. 2018), and set the rotational temperature of HCN molecules to $T_{rot} = 23$ K. We estimate a 3-$\sigma$ upper limit production rate of HCN of $2\times10^{24}$ molecules $s^{-1}$ based on our observations. Changing the rotational temperature of HCN to the values measured in a few cometary comae (e.g., Magee-Sauer et al. 2002, Kobayashi et al. 2010) would change our resulting upper limit production rate by a factor of 2-3. The PSG assumes the Haser model (Haser 1957) to calculate the column density of HCN molecules for an assumed production rate, and then to simulate the expected line strength.

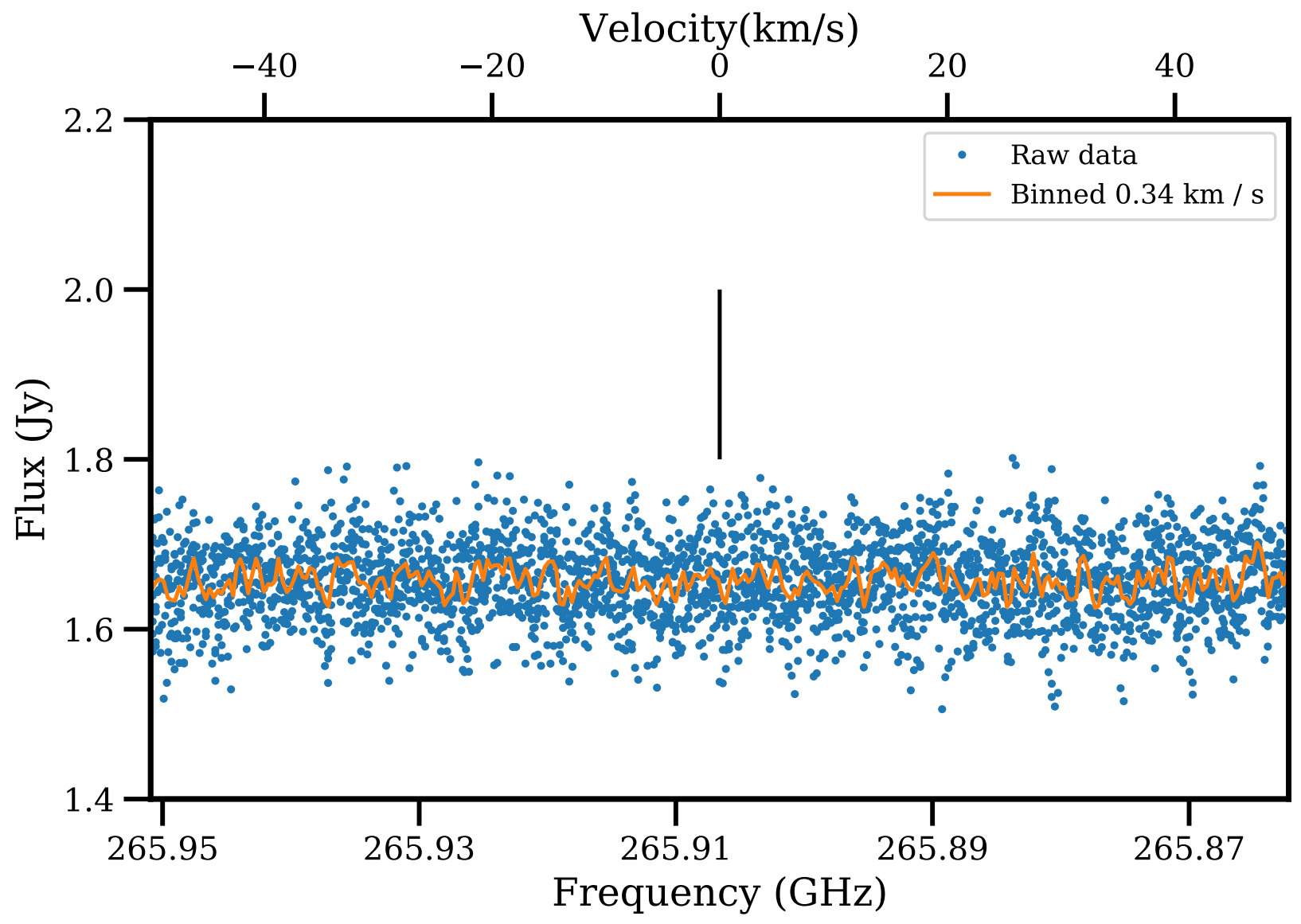

Fig. 3. The spectrum of Ceres centered at the expected frequency of HCN J=3-2 emission line. The vertical line at velocity of 0 km $s^{-1}$ marks the expected frequency of HCN line. No emission line is evident in the smoothed spectrum of Ceres.

4. Thermophysical modeling

In order to understand the thermal lightcurve and measured brightness temperature from our ALMA data, we modeled the disk-averaged brightness temperature of Ceres at 265 GHz at the geometry of our observations in Cycles 4 and 5 (Appendix A). The effect of subsurface emission is determined by the complex dielectric constant of the surface and subsurface material, in particular the refractive index, $n$, and the loss tangent, $\tan\Delta$. The dielectric constant for Ceres's regolith is unknown, and laboratory measurements of permittivity are limited to a small number of minerals and at much lower frequencies (up to tens of GHz) than our data. We can assumed the values based on very limited laboratory measurements and radar observations of the Moon and other solar system objects, and explored the effects of the assumed dielectric constant on our modeled disk-averaged brightness temperature.

[2] https://psg.gsfc.nasa.gov/, accessed on February 25, 2020

Figure 4 shows modeled disk-averaged 265 GHz brightness temperatures plotted as functions of loss tangent for thermal inertia values from 10 to 320 tiu. Based on these modeling results, our brightness temperature measurement of about 173 K from our 2017 September and October observations suggests that the loss tangent is in the range of about 0.1 – 0.3, and the corresponding electrical absorption length is about 0.5 – 2 mm, or 0.5 – 2 wavelengths. The diurnal thermal skin depth is about 1 mm for a thermal inertia of 10 tiu, and proportional to thermal inertia. Therefore, modeling suggests that from our data we are probing within 2 diurnal thermal skin depths or less for any possible thermal inertia of Ceres, and well within the annual thermal skin depth (~7 cm for a thermal inertia of 10 tiu).

The loss tangent inferred from our observations and thermal modeling is much higher than the values measured or estimated for lunar like materials (Gold et al. 1976, Garry & Keihm 1978,

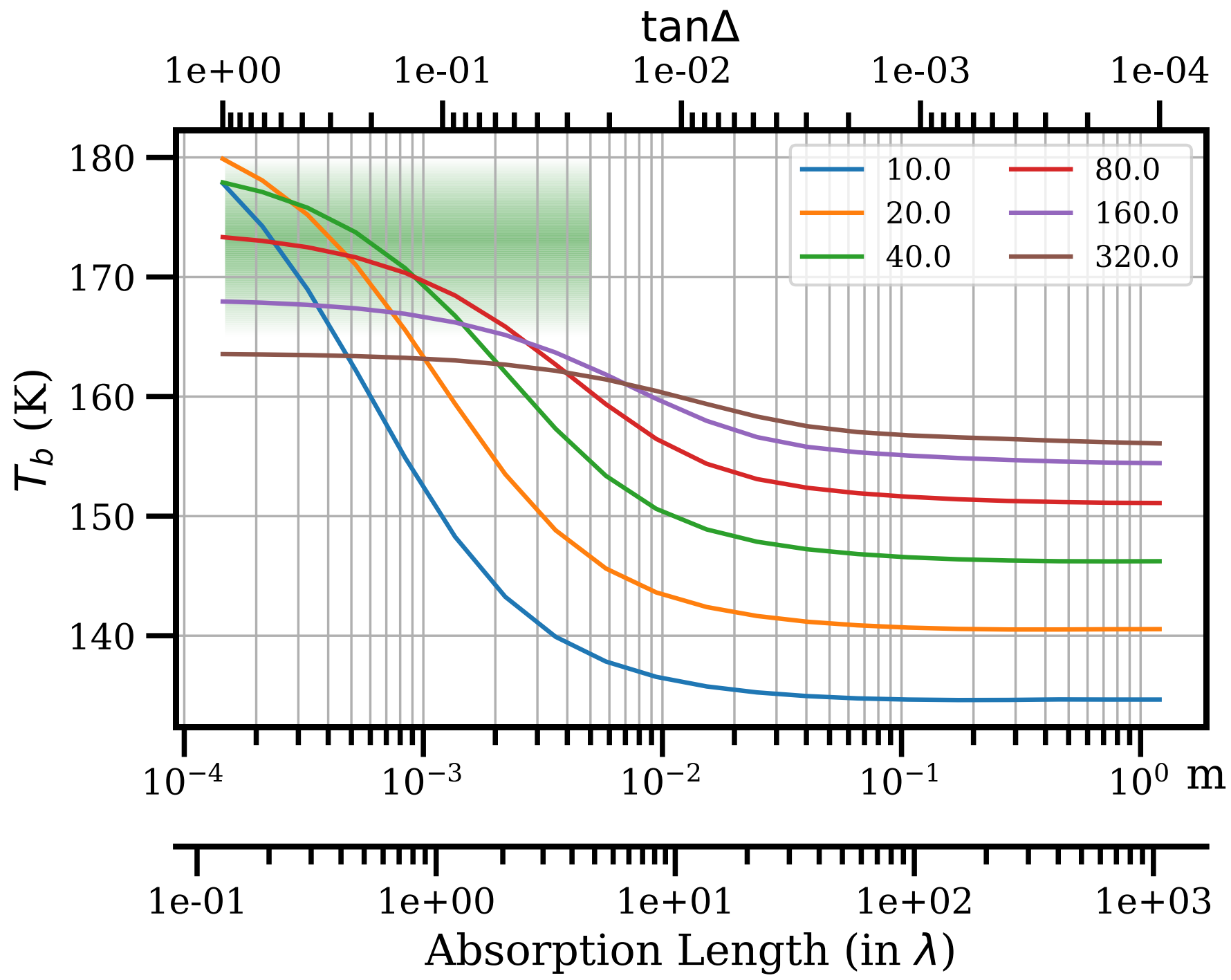

Fig. 4. Modeled disk-averaged brightness temperature plots for Ceres for the geometry of our Cycles 4 and 5 observations with respect to electrical absorption characteristics and thermal inertia. Different curves correspond to different thermal inertia as listed in the figure in units of tiu. The green shaded area marks the range of our measured brightness temperatures centered at about 173 K. The upper x-axis is loss tangent. The two lower x-axes are absorption length in meters and in unit of wavelength (1.13 mm), respectively. The measured range of brightness temperatures allows for a range of loss tangent >~0.02, or electrical absorption length <~5 mm or ~5 wavelengths regardless of thermal inertia.

Keihm & Langseth 1975), comet 67P (Schloerb et al. 2015), as well as for those adopted in all previous similar modeling efforts for Ceres. However, the brightness temperature measured by Chamberlain et al. (2009) at a slightly shorter wavelengths was much lower than ours (130 – 160 K at 345 GHz), that fitted well with their low assumed loss tangent of 0.004. In addition, in the recent effort by Keihm et al. (2013) to simultaneously model the brightness temperatures of Ceres measured from wavelengths of 100 µm to 20 cm, an electrical absorption length of 7 wavelengths was adopted assuming lunar like dielectric properties. But their models resulted in consistently lower modeled brightness temperatures than measurements at mm wavelengths for thermal inertia values between 5 and 200 tiu. Therefore, we find that our inferred high loss tangent is in fact consistent with previous modeling results.

Our modeling also put some constraints on Ceres's thermal inertia. If we assume loss tangent values towards the lower end of the inferred likely range, then the corresponding thermal inertia of Ceres is > 40 tiu. On the other hand, for thermal inertia values higher than 160 tiu, our modeled disk-averaged brightness temperatures would be lower than our measurements for any value of loss tangent. The possible range of thermal inertia of Ceres is therefore probably between 40 and 160 tiu. This range of thermal inertia is also consistent with the indications from previous modeling (Keihm et al. 2013), as well as the new modeling results from Dawn data in the 3 – 5 µm wavelength (Rognini et al. 2019).

# 5. Discussion

## 5.1. Lightcurve and longitudinal thermal variations

What causes the 4% variations in Ceres's thermal flux at a wavelength of 1 mm? Ceres has an oblate shape with a difference of just 2 km between the two equatorial semi-axes (Russell et al. 2016), or 0.4%, too small to explain the observed variation in the thermal lightcurve. The bolometric Bond albedo, $A_B$, of Ceres is about 0.037 (Li et al. 2019). Because the surface temperature is proportional to $(1\text{-}A_B)^{0.25}$, the 4% peak-to-peak range of Ceres's visible lightcurve (Li et al. 2006, Reddy et al. 2015) is only able to explain about a 0.1% of variability in the thermal flux. Therefore, only variations in thermal properties and the physical properties that affect thermal emission at a wavelength of 1 mm can be responsible for the lightcurve variation that we observed.

Our thermophysical modeling shows that, if we allow thermal and dielectric properties to vary, we can easily reproduce the observed thermal lightcurve amplitude. For example, Figure 4 shows that a variation of about 0.1 – 0.4 for the loss tangent will cause the observed brightness temperature to move about 7 – 10 K for a thermal inertia of 40 tiu, enough to explain the observed amplitude of the thermal lightcurve. On the other hand, if the loss tangent is fixed, then the brightness temperature will move along the vertical direction in Figure 4, with the possible range depending on the loss tangent. Only when the loss tangent is higher than about 0.3 could thermal inertia variations cause the observed brightness temperature amplitude. Therefore, it is probably unlikely that thermal inertia variations can fully account for our observed thermal lightcurve amplitude. Of course, this is an extremely simplified model. The reality could be that both thermal inertia and dielectric properties, and maybe also other properties all vary with longitude, and they could also vary with depth, temperature, or both.

Physical properties that affect thermal inertia and dielectric properties include mineralogical composition, albedo, porosity, grain size, distribution and mixing characteristics of different compositions. The geology of Ceres has been thoroughly mapped by the Dawn mission (Williams et al. 2018 and other related papers). Overall, the eastern hemisphere of Ceres is dominated by relatively brighter materials in the Vendimia Planitia region, and contains a relatively high abundance of $NH_4$-phyllosilicates, while the western hemisphere is dominated by fresh, relatively dark ejecta from the Occator crater (Ammannito et al. 2016, Schröder et al. 2017, Li et al. 2019). Therefore, differences in the composition and physical properties of the eastern and western hemispheres could be one explanation for our observed thermal lightcurve. If the $NH_4$-phyllosilicate rich material in the eastern hemisphere is more transparent to mm wavelength radiation, we may be probing deeper into the regolith as a result and therefore measuring lower thermal fluxes. We note that this is just one possible explanation for the thermal lightcurve, and other possibilities definitely exist.

### 5.2. Implications of high electrical absorption

Based on our modeling assumptions (single-layer thermal model, constant thermal parameters as listed in Table A1, and constant dielectric constant), the electrical absorption of Ceres's surface is much higher than that of the lunar surface. It may also be higher than that of the surface of 67P and many other objects observed with radar because assumed lunar-like dielectric characteristics have been sufficient for modeling those observations. This characteristic has also been noticed by Redman et al. (1998), who concluded that the outermost surface layer of Ceres "must be partially, but not completely, opaque at wavelengths near 1 mm".

Different thermal emission behavior in the 1 mm wavelength region have been observed for other asteroids. Based on modeling efforts similar to what they performed for Ceres, Keihm et al. (2013) showed that Vesta, the prototype of V-type asteroids, appears to be much cooler than the model predictions (unlike Ceres) in this range of wavelengths (see their Figs. 5 & 6). Meanwhile, they found that Pallas, a B-type asteroid, shows an overall good fit to model predictions from 10 μm to cm wavelengths (see their Fig. 8). Interestingly, Hygeia, which has a similar taxonomic type as Ceres, shows similar behavior as Ceres with hotter observed brightness temperatures than the modeled values in this wavelength range, although it only has two measurements in the sub-mm and mm wavelength region (see their Fig. 9). In addition, Redman et al. (1998) analyzed the thermal emission of various asteroids with respect to wavelength based on their effective emissivities with respect to the predictions by the STM, and their results for Ceres and Vesta are entirely consistent with the Keihm et al. (2013) modeling results. On the other hand, Redman et al. (1998) reported that (6) Hebe and (18) Melpomene, both S-type, show similar behavior as Vesta, while M-type asteroids (16) Psyche and (216) Kleopatra show even cooler brightness temperatures than those of Vesta and S-type asteroids compared to the STM predictions in the 1 mm wavelength region. Interestingly, the S-type asteroid (7) Iris also displays similar behavior as Ceres and Hygeia at mm wavelengths (Redman et al. 1998).

Given the above results, it is reasonable to suspect that brightness temperatures at sub-mm to mm wavelengths are correlated with taxonomic type, which corresponds to the type of materials in the thermally active layers on an asteroid's surface. Vesta has a basaltic surface (Binzel & Xu 1993, De Sanctis et al. 2012), and therefore behaves similarly to the Moon. S-type asteroids contains similar silicate minerals on the surface with similar density as Vesta and the Moon. On the other hand, the surfaces of Ceres, Pallas, and Hygeia are dominated by carbonaceous materials and hydrated minerals (c.f. Rivkin et al. 2015). Ceres and Hygeia have similar reflectance spectral shapes in the 3 µm region with a distinct absorption feature at ~3.05 µm superimposed on a broader absorption feature from 2.8 to 3.7 µm, whereas Pallas displays a sharp feature (Takir & Emery 2012, Rivkin et al. 2012). It is possible that the unique thermal emission behavior at a wavelength of 1 mm for Ceres is associated with the distinct composition of its top layer regolith.

The composition of Ceres's surface is dominated by ammoniated phyllosilicates that cause spectral features at 2.7 µm and 3.05 µm (De Sanctis et al. 2015, Ammannito et al. 2016). Carbonates and chloride salts are enriched in local areas of Ceres's surface such as the Occator crater (e.g., Raponi et al. 2019), and water ice is exposed in some isolated regions of up to a few $km^2$ in area on crater walls (Combe et al. 2016). Dominating the shallow subsurface of Ceres is a mixture of rocks, salts and/or clathrates accounting for 60% to 70% of the material by volume (Bland et al. 2016). Ceres's surface could also be rich in graphitized carbon (Hendrix et al. 2018). Laboratory measurements show relatively higher loss tangent values for phyllosilicates than for anhydrate silicates in general, and even higher loss tangent values for various phases of carbons (cf. Herique et al. 2016, 2018), although those measurements were all made at dm to m wavelengths, which are much longer than those of our observations. Therefore, the hot surface of Ceres at sub-mm and mm wavelengths could be due to the unique composition of Ceres's top layer, where regolith material is opaque to mm wavelength thermal emission, but more transparent to longer wavelength due to material with grain sizes of tens of µm or smaller that makes up the top surface of Ceres. The indication of abundant small grains in Ceres regolith is also consistent with the photometric study in the optical wavelength (Li et al. 2019) and thermal analysis (Schorghofer 2016). In addition, the similar SED characteristics of Hygiea's thermal emission suggests that both it and Ceres possibly share similar surface mineralogical composition and physical properties.

## 6. Conclusions

The disk-integrated thermal emission of Ceres as observed by ALMA at a wavelength of 1.13 mm shows a rotational lightcurve with a variation of about 4% (peak-to-peak) and a disk-averaged brightness temperature of 170 – 180 K. A search for HCN around Ceres resulted in a non-detection with an estimated upper limit production rate of $2\times10^{24}$ molecules $s^{-1}$, assuming a globally uniform production and a Haser model.

Thermophysical modeling assuming uniform thermal and dielectric parameters suggests that Ceres's top layer has a loss tangent of about 0.1 – 0.3, corresponding to an electric absorption length of about one thermal skin depth at mm wavelengths. This dielectric absorption is much stronger than that of lunar-like materials at similar wavelengths, and consistent with previous observations. On the other hand, previous observations at longer wavelengths are consistent with relatively lower loss tangents, suggesting that the particles covering the top surface of Ceres are relatively less transparent at a wavelength of 1 mm than at other wavelengths. Comparing Ceres with other large asteroids and the Moon, we suggest that the high dielectric absorption that we observed at mm-wavelengths is related to Ceres's unique surface composition, which is dominated by hydrated minerals and high abundances of carbonates and salt, as well as the ~µm-sized grains covering the surface. Thermophysical analysis also suggests that observed thermal lightcurve variations are likely dominated by spatial variations in the dielectric properties on Ceres's surface, which may be related to the compositional variations, specifically in the abundance of phyllosilicates.

The seasonal 1-mm wavelength brightness temperature variability of Ceres possibly shows a large lag with respect to the mean surface temperature predictions by STM based on the heliocentric distance of Ceres. Future observations are needed to better define the seasonal temperature trend and understand the nature and cause of the possible thermal lag.

## Acknowledgements

This research is supported by NASA's Solar System Observations (SSO) R&A Program through Grant NNX15AE02G to the Planetary Science Institute. JYL and HHH also acknowledge partial support from the Solar System Exploration Research Virtual Institute 2016 (SSERVI16) Cooperative Agreement (Grant NNH16ZDA001N), SSERVI-TREX to the Planetary Science Institute. This paper makes use of the following ALMA data: ADS/JAO.ALMA#2015.1.01384.S, ADS/JAO.ALMA#2016.1.00748.S, ADS/JAO.ALMA#2017.1.0640.S. ALMA is a partnership of ESO (representing its member states), NSF (USA) and NINS (Japan), together with NRC (Canada), MOST and ASIAA (Taiwan), and KASI (Republic of Korea), in cooperation with the Republic of Chile. The Joint ALMA Observatory is operated by ESO, AUI/NRAO and NAOJ. The National Radio Astronomy Observatory is a facility of the National Science Foundation operated under cooperative agreement by Associated Universities, Inc. JYL is grateful to Dr. Geronimo Villanueva for his help on the PSG and upper limit estimate of HCN, and to Prof. Wenzhe Fa（法文哲）and Dr. Ladislav Rezac for helpful discussions in the thermophysical modeling and interpretations of the observational results. This research made use of a number of open source Python packages (in arbitrary order): Astropy (Astropy Collaboration 2013), Matplotlib (Hunter 2007), SciPy (Jones et al. 2001), IPython (Pérez and Granger 2007), Jupyter notebooks (Kluyver et al. 2016).

## Appendix A

The modeling of Ceres's disk-averaged brightness temperature performed in this work is a two-step process. In the first step, we performed thermophysical modeling to calculate the surface and subsurface temperature profiles and distributions using the USGS KRC code (Kieffer 2013). We used a triaxial ellipsoidal shape model of Ceres (Russell et al. 2016) and geometric parameters corresponding to our observations in Cycles 4 and 5. Table A1 lists the thermal parameters that we used for Ceres. We did not include the effects of surface roughness and infrared beaming in our thermophysical modeling. This is because thermal emission at sub-mm and mm wavelengths is dominated by subsurface emission, while roughness and thermal beaming effects primarily affect the temperature distribution of the top surface (Keihm et al. 2013).

Once temperature distributions on the disk of Ceres are calculated for depths down to 14 thermal skin depths, below which there is almost no temperature fluctuation, we use radiative transfer theory (e.g., Hapke 2012) to calculate observed temperature distributions in order to account for thermal emissions from the subsurface. Under Rayleigh-Jean's approximation at mm wavelengths, the measured brightness temperature is the integral of the subsurface temperature profile as a function of depth, *T*(*z*), using,

$$T_b = (1 - R_\nu)\int T(z)\exp\left(-\frac{z}{L_e\cos i}\right)\frac{dz}{L_e\cos i}$$

where *i* is the incidence angle of thermal emission before it crosses the surface boundary, which is related to emission angle *e* by Snell's refraction law, and depends on the refractive index, *n*. $R_\nu$ is the Fresnel reflection coefficient, which depends on *i* and *n*, and in our case, is the average of the reflection coefficients of both polarization states. $L_e$ is the electrical skin depth (or electrical absorption length), which is the reciprocal of absorption coefficient, $\kappa$, which is given by,

$$\kappa = \frac{4\pi n}{\lambda}\sqrt{\frac{\sqrt{(1+\tan^2\Delta)}-1}{2}}$$

where $\lambda$ is wavelength and $\tan\Delta = K_{ei}/K_{er}$ is the loss tangent, where $K_{er}$ and $K_{ei}$ are the real and imaginary components of the complex dielectric constant, $K_e$, respectively. The refractive index, *n*, is the square root of the real part of complex dielectric constant, $n = \sqrt{K_{er}}$. With the radiative transfer model, we can derive the observed brightness temperature distribution from the thermophysical model and calculate the disk-averaged brightness temperature.

Brouet et al. (2015) measured the relative permittivity (the real part of the dielectric constant) of dry porous regolith analogs at frequencies from 50 MHz to 90 GHz in the laboratory to be in the range of 1.5 to 2.2, with slight variations with frequency. When the mean grain sizes are equivalent to <10 wavelengths in diameter, the dielectric constant is nearly independent of wavelength. For grain sizes close to one wavelength in diameter, the permittivity increases towards >3. Brouet et al. (2016) measured the permittivity of a water ice-dust mixture from 5 MHz to 2 GHz, and found it to be 1.1 to 2.7, almost independent of frequency. Porosity will decrease the dielectric constant. The top layer of the Ceres regolith is dominated by hydrated minerals (De Sanctis et al. 2015, Ammannito et al. 2016), although water ice is identified in a number of isolated areas of km in size (Combe et al. 2016) and could also exist in permanently shadowed regions near the poles (Platz et al. 2016). Assuming a porosity of 0.5 for Ceres surface material, the real part of the dielectric constant for Ceres is then likely between 1.9 and 2.7, and the corresponding refractive index is 1.4 to 1.6. We adopted a value of 1.5 in our modeling, close to the lunar like real part of the dielectric constant value of 2.3 (Gold et al. 1976). Modeling suggests that the resulting brightness temperature varies by less than 8 K with refractive index varying between 1.9 and 2.7 for a loss tangent of 0.1.

On the other hand, the measurements of loss tangent are much more limited, especially in the laboratory at frequencies of hundreds of GHz. Lunar surface material has been measured from the ground previously. Garry and Keihm (1978) reported that the absorption length of lunar regolith is equivalent to about 7× wavelengths in mm bands and 10-15 wavelengths in cm bands. Work by Keihm & Langseth (1975) suggested that lunar-like regolith had loss tangents of 0.04 and 0.002 at wavelengths of 0.1 cm and 68 cm, respectively. Rosetta/MIRO observations of Comet 67P/Churyumov-Gerasimenko suggested that the comet had an electrical penetration depth of 3.9 cm at a wavelength of 1.5 mm and 1 cm at a wavelength of 0.5 mm (Schloerb et al. 2015). In the modeling of MIRO observations of Asteroid Lutetia, Gulkis et al. (2012) adopted values of 0.019 and 0.014 for the loss tangent at wavelengths of 0.5 mm and 1.5 mm, respectively, based on lunar like materials measured from the ground by Gold et al. (1976). Chamberlain et al. (2009) adopted a value of 0.0040 for the loss tangent of Ceres when modeling their observations at 345 GHz, based on the empirical relationship between loss tangent and surface density as proposed by Ostro et al. (1999), although they suggested a possibly higher loss tangent from their results. Therefore, 0.004 to 0.04 is likely a reasonable range of values to use for the loss tangent of Ceres, and the corresponding absorption lengths are on a few wavelengths. We experimented with values between $10^{-4}$ and 1 to suggest the range of likely values for Ceres.

**References**

A'Hearn, M.F., & Feldman, P.D., 1992, Icar, 98, 54
Altenhoff, W.J., Johnston, K.J., Stumpff, P., et al., 1994, A&A, 287, 641
Altenhoff, W.J., Baars, J.W.M., Schraml, J.B., et al., 1996, A&A, 309, 953
Ammannito, E., De Sanctis, M.C., Ciarniello, M., et al., 2016. Science 353, 1006
Astropy Collaboration, 2013, A&A, 558, A33
Binzel, R.P., & Xu, S., 1993, Sci, 260, 186
Bland, M.T., Raymond, C.A., Schenk, P.M., et al. 2016. NatGeo, 9, 538
Brouet, Y., Levasseur-Regourd, A.C., Sabouroux, P., et al., 2015, A&A, 583, A39
Brouet, Y., Neves, L., Sabouroux, P., et al., 2016, JGRE, 121, 2426
Chamberlain, M.A., Sykes, M.V., & Esquerdo, G.A., 2007, Icar, 188, 451
Chamberlain, M.A., Lovell, A.J., & Sykes, M.V., 2009, Icar, 202, 487
Combe, J.-P., McCord, T.B., Federico, T., et al., 2016, Sci, 353, 1007
Combe, J.-P., Raponi, A., Federico, T., et al., 2019, Icar, 318, 22
Cordiner, M.A., Remijan, A.J., Boissier, J., et al., 2014, ApJL, 792, 2
De Sanctis, M.C., Ammannito, E., Capria, M.T., et al., 2012, Sci, 336, 697
De Sanctis, M.C., Ammannito, E., Raponi, A., et al., 2015, Natur, 528, 241
De Sanctis, M.C., Raponi, A., Ammannito, E., et al., 2016, Natur, 536, 54
Gulkis, S. Keihm, S., Kamp, L., et al., 2012, P&SS, 66, 31
Gary, B.L., & Keihm, S. J., 1978, LPSC 9 (A79-39253 16–91), 3, 2885
Gold, T., Bilson, E., & Baron, R.L., 1976, LPSC 7 (A77-34651 15–91), 3, 2593
Hapke, B., 2012, Theory of reflectance and emittance spectroscopy (2nd ed.; Cambridge University Press)
Hendrix, A.R., Vilas, F., & Li, J.-Y., 2016, GeoRL, 43, 8920
Hendrix, A.R., Hurford, T.A., Barge, L.M., et al., 2019, Astrobio, 19, 1
Herique, A., Kofman, W., Beck, P., et al., 2016, MNRAS, 462, 516
Herique, A., Agnus, B., Asphaug, E., et al., 2018, AdSpR, 62, 2141
Hiesinger, H., Marchi, S., Schmedemann, N., et al., 2016, Sci, 353, id.aaf4758
Hunter, J.D., 2007, CSE, 9, 90
Jones, E., et al., 2001, SciPy: Open Source Scientific Tools for Python, http://www.scipy.org.
Keihm, S.J., & Langseth, M.G., 1975, Icar, 24, 211
Keihm, S., Kamp, L., Gulkis, S., et al., 2013, Icar, 226, 1086
Kieffer, H.H., 2013, JGRE, 118, 451
Kluyver, T., Ragan-Kelley, B., Pérez, F., et al., 2016, https://doi.org/10.3233/978-1-61499-649-1-87
Kobayashi, H., Bockelée-Morvan, D., Kawakita, H., et al., 2010, A&A 509, 80
Küppers, M., O'Rourke, L., Bockelée-Morvan, D., et al., 2014, Natur, 505, 525
Lebofsky, L.A., Sykes, M.V., Tedesco, E.F., et al., 1986, Icar, 68, 239
Li, J.-Y., McFadden, L.A., Parker, J.Wm., et al., 2006, Icar, 182, 143
Li, J.-Y., Reddy, V., Nathues, A., et al. 2016, ApJL, 817, 22
Li, J.-Y., Schröder, S.E., Mottola, S., et al., 2019, Icar, 322, 144
Magee-Sauer, K., Mumma, M.J., DiSanti, M.A., et al., 2002, JGR 107, E11, 5096

McMullin, J.P., Waters, B., Schiebel, D., Young, & M., Golap, K., 2017, ASP Conference Ser. 376, proceedings, 127.
Mitchell, D.L., Ostro, S.J., Hudson, R.S., et al., 1996, Icar, 124, 113
Moullet, A., Gurwell, M., & Carry, B., 2010, A&A Lett., 516, 10
Ostro, S.J., Hudson, R.S., Rosema, K.D., et al., 1999, Icar, 137, 122
Park, R.S., Konopliv, A.S., Bills, B.G., et al., 2016, Natur, 537, 515
Pérez, F., & Granger, B.E., 2007, CSE, 13, 22
Platz, T., Nathues, A., Schorghofer, N., et al., 2016, NatAs, 1, 7
Prettyman, T.H., Yamashita, N., Toplis, M.J., et al., 2017, Sci, 355, 55
Raponi, A., De Sanctis, M.C., Carrozzo, F.G., et al., 2019, Icar, 320, 83
Reddy, V., Li, J.-Y., Gary, B.L., et al., 2015, Icar, 260, 332
Redman, R.O., Feldman, P.A., & Mathews, H.E., 1998, AJ, 116, 1478
Rivkin, A. S., Howell E. S., Emery J. P., et al., 2012, EPSC 2012, 359.
Rivkin, A.S., Campins, H., Emery, J.P., et al., 2015, in Asteroids IV, ed. P. Michel et al. (Tucson, AZ: Arizona Univ. Press), 65
Rognini, E., Capria, M.T., Tosi, F., et al., 2019, JGRE, in press. doi: 10.1029/2018JE005733.
Ruesch, O., Platz, T., Schenk, P., et al. 2016, Sci, 353, id.aaf4286
Ruesch, O., Quick, L.C., Landis, M.E., et al. 2019, Icar, 320, 39
Russell, C.T., Raymond, C.A., Ammannito, E., et al., 2016, Sci, 353, 1008
Saint-Pé, O., Combes, M., & Rigaut, F., 1993, Icar, 105, 271
Schloerb, F.P., Keihm, S., von Allmen, P., et al., 2015, A&A, 583, A29
Schorghofer, N., 2008, ApJ, 682, 697
Schorghofer, N., 2016, Icar, 276, 88
Schorghofer, N., Mazarico, E., Platz, T., et al., 2016, GeoRL, 43, 6783
Schröder, S.E., Mottola, S., Carsenty, U., et al. 2017, Icar, 288, 201
Scully, J.E.C., Bowling, T., Bu, C., et al., 2019, Icar, 320, 213
Sizemore, H.G., Platz, T., Schorghofer, N., et al., 2017, GeoRL, 44, 6570
Sori, M.M., Sizemore, H.G., Byrne, S., et al. 2018, NatAs, 2, 946
Spencer, J.R., 1990, Icar, 83, 27
Takir, D., & Emery, J.P., 2012, Icar, 219, 641
Titus, T.N., 2015, GeoRL, 42, 2130
Ulich, B.L., Dickel, J.R., & de Pater, I., 1984, Icar, 60, 590
Villanueva, G.L., Smith, M.D., Protopapa, S., et al., 2018, JQSRT, 217, 86
Webster, W.J., Jr., Johnston, K.J., Hobbs, R.W., et al., 1988, AJ, 95, 1263
Webster, W.J., & Johnston, K.J., 1989, PASP, 101, 122
Williams, D.A., Buczkowski, D.L., Mest, S.C., et al., 2018, Icar, 316, 1

Table 1 Dates and geometric circumstances of our ALMA observations of Ceres, and the measured total flux density and brightness temperature. The uncertainties listed here are the standard deviations of measurements from each date. The photon counting error is much smaller than the scatter of measurements.

| ALMA Cycle | Date | UT Start | UT Stop | $r_h$ (au) | $\Delta$ (au) | Local Solar Time | Sub-Earth Longitude (deg) | Flux Density (Jy) | $T_B$ (K) |
|---|---|---|---|---|---|---|---|---|---|
| 3 (12-m) | 2015-10-31 | 22:59:46 | 02:08:46 | 2.97 | 2.929 | 13:16 | 356 – 240 | 1.06±0.03 | 138±5 |
| | 2015-11-01 | 23:00:05 | 02:07:27 | | 2.943 | | 124 – 0 | 0.99±0.06 | 133±8 |
| | 2015-11-12 | 19:32:01 | 20:57:24 | | 3.092 | | 233 – 184 | 0.61±0.10 | 83±14 |
| | | 22:20:04 | 23:36:26 | | 3.092 | | 128 – 79 | 0.84±0.11 | 114±15 |
| 4 (12-m) | 2017-09-26 | 11:09:29 | 13:02:54 | 2.63 | 2.932 | 10:40 | 348 – 278 | 1.30±0.01 | 172±2 |
| | 2017-09-27 | 11:16:25 | 13:08:38 | | 2.919 | | 108 – 42 | 1.29±0.02 | 169±3 |
| | 2017-09-28 | 10:55:22 | 12:01:12 | | 2.907 | | 254 – 215 | 1.33±0.02 | 173±3 |
| | 2017-09-30 | 10:50:35 | 11:43:48 | | 2.882 | | 153 – 124 | 1.34±0.02 | 171±3 |
| 5 (12-m) | 2017-10-15 | 10:50:54 | 13:15:46 | 2.62 | 2.688 | 10:33 | 267 – 191 | 1.35±0.10 | 151±10 |
| | 2017-10-19 | 10:43:15 | 12:57:02 | | 2.634 | | 68 – 352 | 1.61±0.03 | 172±3 |
| | 2017-10-26 | 11:06:34 | 12:06:05 | | 2.540 | | 232 – 202 | 1.67±0.03 | 166±3 |
| 5 (ACA) | 2017-10-15 | 12:29:06 | 13:29:34 | 2.62 | 2.694 | 10:33 | 212 – 172 | 1.4±0.2 | 163±10 |
| | 2017-10-18 | 09:28:10 | 11:56:08 | | 2.654 | | 354 – 258 | 1.60±0.05 | 181±5 |
| | 2017-10-19 | 09:24:44 | 11:52:28 | | 2.641 | | 126 – 29 | 1.60±0.05 | 179±5 |
| | 2017-10-20 | 10:45:50 | 11:22:36 | | 2.627 | | 201 – 177 | 1.61±0.05 | 178±5 |

Table 2. Historical measurements of the total flux and brightness temperature at frequencies close to our data (265 GHz) in the literature, including those from ALMA archive.

| Date | Frequency (GHz) | $r_h$ (au) | $\Delta$ (au) | Local Solar Time | True Anomaly (deg) | Flux (Jy) | Tb (K) | Facility | Reference |
|---|---|---|---|---|---|---|---|---|---|
| 2015-08-31 | 230.6 | 2.957 | 2.135 | 12:53 | 156 | 2.02±0.05 | 187±5 | ALMA | ALMA Archive |
| 2015-08-27 | 239.7 | 2.956 | 2.099 | 12:49 | 155 | 1.92±0.05 | 159±4 | ALMA | ALMA Archive |
| 2015-08-14 | 227.0 | 2.952 | 2.004 | 12:33 | 153 | 1.97±0.05 | 166±4 | ALMA | ALMA Archive |
| 2015-08-14 | 264.3 | 2.952 | 2.004 | 12:33 | 153 | 2.75±0.07 | 171±1 | ALMA | ALMA Archive |
| 2009-01-28 | 235.0 | 2.547 | 1.692 | 11:05 | 357 | 3.30±0.02 | 185±1 | SMA | Moullet et al. (2010) |
| 1995-05-23 | 264.0 | 2.559 | 2.590 | 13:30 | 8 | 1.79±0.11 | 186±12 | JCMT | Redman et al. (1998) |
| 1995-04-10 | 250.0 | 2.557 | 2.182 | 13:27 | 10 | 2.36±0.02 | 194±2 | HHT | Altenhoff et al. (1996) |
| 1993-07-17 | 264.0 | 2.927 | 2.857 | 10:39 | 218 | 1.44±0.07 | 182±8 | JCMT | Redman et al. (1998) |
| 1993-07-17 | 233.0 | 2.927 | 2.857 | 10:39 | 218 | 1.13±0.08 | 184±12 | JCMT | Redman et al. (1998) |
| 1989-03-10 | 250.0 | 2.878 | 3.700 | 12:39 | 235 | 0.75±0.03 | 177±8 | IRAM | Altenhoff et al. (1994) |
| 1987-12-27 | 250.0 | 2.955 | 3.831 | 12:30 | 153 | 0.54±0.08 | 137±19 | IRAM | Altenhoff et al. (1994) |
| 1983-05-24 | 227.0 | 2.954 | 2.662 | 10:40 | 152 | 1.11±0.14 | 165±21 | KP12m | Ulich et al. (1984) |
| 1983-05-19 | 227.0 | 2.953 | 2.727 | 10:39 | 151 | 1.11±0.18 | 173±28 | KP12m | Webster et al. (1988) |

Table A1. Thermal parameters adopted in our modeling

| Parameter | Value | Source and/or Reference |
|---|---|---|
| Triaxial ellipsoidal shape | 482.64 × 480.64 × 445.57 km | Russell et al., (2016) |
| Pole (RA, Dec) | 291.421º, 66.758º | Park et al. (2016) |
| Subsolar latitude | -3.8º | ALMA Cycle 4 & 5 |
| Sub-Earth latitude | -0.7º | ALMA Cycle 4 & 5 |
| Heliocentric distance | 2.63 au | ALMA Cycle 4 & 5 |
| Bolometric Bond albedo | 0.037 | Li et al. (2019) |
| Rotational period | 9.074170 hours | Chamberlain et al. (2007) |
| Emissivity | 0.95 | Keihm et al. (2013) |
| Surface density | $1.24\times10^{3}$ kg $m^{-3}$ | Michell et al. (1996) |
| Specific heat | 750 J $(K\ kg)^{-1}$ | Typical rock value |